\documentclass[12pt,preprint]{emulateapj}

\usepackage{natbib}
\def\degree{\ifmmode {^\circ}\else {$^\circ$}\fi}
\def\rstar{\ifmmode {\, R_{\star}}\else $R_{\star}$\fi}
\def\msol{\ifmmode {\, M_{\odot}}\else $M_{\odot}$\fi}
\def\rsol{\ifmmode {\, R_{\odot}}\else $R_{\odot}$\fi}
\def\lsol{\ifmmode {\, L_{\odot}}\else $L_{\odot}$\fi}
\def\msolyr{\ifmmode {\, M_{\odot}\,{\rm yr}^{-1}}\else $M_{\odot}\,{\rm yr}^{-1}$\fi}
\def\mdot{\ifmmode {\,\dot{M}}\else $\dot{M}$\fi}
\def\mdotyr{\ifmmode {\,\dot{M}\,yr^{-1}}\else $\dot{M}\,yr^{-1}$\fi}

\newcommand{\Teff}{\ifmmode{T_{\rm eff}}\else{$T_{\rm eff}$}}

\begin{document}

\title{Dusty Disks Around Central Stars of Planetary Nebulae}

\author{Geoffrey C. Clayton\altaffilmark{1}, Orsola De Marco\altaffilmark{2,3},  Jason Nordhaus\altaffilmark{4}, Joel Green\altaffilmark{5}, Thomas Rauch\altaffilmark{6}, Klaus Werner\altaffilmark{6}, and You-Hua Chu\altaffilmark{7}}

\altaffiltext{1}{Department of Physics \& Astronomy, Louisiana State
University, Baton Rouge, LA 70803; gclayton@fenway.phys.lsu.edu}
\altaffiltext{2}{Department of Physics \& Astronomy, Macquarie University, Sydney, NSW 2109, Australia; orsola@science.mq.edu.au}
\altaffiltext{3}{Astronomy, Astrophysics, and Astrophotonics Research Centre, Macquarie University, Sydney, NSW 2109, Australia}
\altaffiltext{4}{NSF Fellow, Center for Computational Relativity and Gravitation, and National Technical Institute for the Deaf, Rochester Institute of Technology, Rochester, NY 14623; : nordhaus@astro.rit.edu}
\altaffiltext{5}{The University of Texas, Department of Astronomy, 1 University Station, C1400, Austin, TX 78712-0259; joel@astro.as.utexas.edu}
\altaffiltext{6}{Institute for Astronomy \& Astrophysics, Kepler Center for Astro and Particle Physics, Eberhard Karls University, Sand 1, 72076, T\"{u}bingen, Germany; rauch, werner@astro.uni-tuebingen.de}
\altaffiltext{7}{Department of Astronomy, University of Illinois at Urbana-Champaign, 1002 West Green Street, Urbana, IL 61801; chu@astro.uiuc.edu}

\begin{abstract}

Only a few percent of cool, old white dwarfs (WDs) have infrared excesses interpreted as originating in small hot disks due to the infall and destruction of single asteroids that come within the star's Roche limit. Infrared excesses at 24 \micron~were also found to derive from the immediate vicinity of younger, hot WDs, most of which are still central stars of planetary nebulae (CSPN). The incidence of CSPN with this excess is 18\%. 
The Helix CSPN, with a 24 $\mu$m excess, has been suggested to have a disk formed from collisions of Kuiper belt-like objects (KBOs).
In this paper, we have analyzed an additional sample of CSPN to look for similar infrared excesses. These CSPN are all members of the PG 1159 class and were chosen because their immediate progenitors are known to often have dusty environments consistent with large dusty disks. We find that, overall, PG 1159 stars do not present such disks more often than other CSPN, although the statistics (5 objects) are poor. 
We then consider the entire sample of CSPN with infrared excesses, and compare it to the infrared properties of old WDs, as well as cooler post-AGB stars. We conclude with the suggestion that the infrared properties of CSPN more plausibly derive from AGB-formed disks rather than disks formed via the collision of KBOs, although the latter scenario cannot be ruled out. We finally remark that there seems to be an association between CSPN with a 24-$\mu$m excess and confirmed or possible binarity of the central star. 

\end{abstract}


\keywords{dust --- circumstellar matter --- white dwarfs --- planetary nebulae --- stars: evolution}

\section{Introduction}

There are three indicators of planetary debris around white dwarfs (WDs): (1) metal pollution, discovered decades ago \citep[e.g.,][]{1960ApJ...131..638W}, but originally thought to be accretion from the ISM \citep[e.g.,][]{1993AJ....105.1033A, 1993ApJS...84...73D}, (2) an infrared (IR) excess due to warm dust and first detected around G29-38 \citep{1987Natur.330..138Z}, (3) gaseous disks \citep[e.g.,][]{2006Sci...314.1908G}, which unambiguously showed that the debris material is an accretion disk lying within the tidal disruption radius of the WD \citep{2005ApJ...635L.161R}.
The debris material comes from the disruption of asteroids \citep{1990ApJ...357..216G,2003ApJ...584L..91J}.
These three indicators of debris accretion are hierarchical. All WDs that have gaseous disks also have dust/IR excesses, and all WDs that have dust also have metal-polluted atmospheres \citep{2009ApJ...696.1402B, 2012ApJ...750...86B}. However, this does not work in the opposite direction, i.e., not all metal-polluted WDs have dust, and not all dusty WDs have gas. The fraction of cool WDs with IR excesses is $\sim$1--3\% \citep{2009ApJ...694..805F,2011MNRAS.416.2768S,2011MNRAS.417.1210G,2012ApJ...760...26B}.
These disks are typically found very close ($<$1 R$_{\sun}$) to cool ($<$25,000 K) WDs. The dust in these disks is quite warm ($\sim$1000 K), and so the dust emission shows up strongly in the Spitzer IRAC bands \citep{2012ApJ...745...88X}. 
Many questions still remain about the structure and longevity of these disks and what they are telling us about the planetary systems that once orbited these stars, but other than that, the debris nature of these disks seems to be a reasonable interpretation.

In 2007, a dust disk was detected around the central star of the Helix planetary nebula (PN), also known as NGC~7293 \citep{2007ApJ...657L..41S}. This disk differed greatly from those previously found around WDs. In the Helix system, the star is much hotter (110,000 K vs. $<$25,000 K of the typical WD with debris disks); the dust is much colder ($\sim$100 K vs. 1000 K for the WD disks), and lies much farther from the star \citep[$\sim$50 AU vs. $<$0.01 AU for the WD disks;][]{2007ApJ...657L..41S,2012ApJS..200....3B}. Nonetheless, this object was also interpreted as having a debris disk, but the suggested origin was dust production by collisions among Kuiper belt-like objects (KBOs), rather than individual asteroids pulverized by entering the Roche limit.

Further surveys have found cold dust disks around a number of hot WDs and central stars of PN (CSPN). \citet{2011AJ....142...75C} looked at seventy-one stars and found nine disks. Seven of these nine stars with disks were CSPN. The other two are hot WDs likely to have been PN in the very recent past \citep{1994ApJ...433L..93T,1997A&A...327..721W}. Out of the 35 CSPN analyzed by \citet{2011AJ....142...75C}, 7 detections represent a detection rate of 20\%. \citet{2012ApJS..200....3B} searched an additional sample of CSPN. Out of 56 viable candidates, they detected disks in 17-20\% of their sample. For the combined samples the incidence of disks is 18\%.
The much higher frequency of disks around CSPN than around cool WDs raised questions as to whether their nature is also that of debris disks or whether these disks are formed 
by mass loss from the stars during the asymptotic giant branch (AGB) phase.
Dusty outflows, disks, and shells are, of course, known and expected to be associated with cool AGB and post-AGB stars, but  one might expect them to be destroyed by the heating of their central stars. The presence and longevity of disks around CSPN and young, hot WDs can inform us about the recent past of these stars, including the presence of a binary companion that may have influenced the AGB mass loss process, and also about dust formation and survival.

These questions led us to look for more disks around CSPN. We selected the PG 1159 stars that constitute about 10-20\% of the entire CSPN group. These stars are intermediate between the the Wolf-Rayet central stars of PN (also known as the [WC] stars) and WDs.
The reason for targeting these stars is that some of their immediate progenitors, the [WC] stars, are known to have spectacular, dusty environments, with large silicate disks as well as carbon rich outflows \citep[e.g., CPD -56\degree 8032,][]{1999ApJ...513L.135C}\footnote{All known examples of oxygen-and-carbon chemistry in [WC] stars are in the [WCL] (`L' for late) class, in the immediate post-AGB phase, with temperatures of 30-50kK. Although there have been claims that at least one of the hotter, earlier [WC] objects also shows the dual-dust signature of a disk, we note that the only claim of dual-dust for a [WCE] central star (NGC 5315) is by \citet{2002PASP..114..602D}, who cited a private communication. A similar claim was made by \citet{2009A&A...495L...5P} with no citation. We examined the ISO SWS spectrum of NGC 5315 and there is no sign of silicate features. The ISO spectrum was also classified as 4/5.Eu: by \citet{2004ApJS..151..299H}, which indicates atomic and PAH emission but no silicate features.}. 
Abundance measurements of debris around cool metal-polluted WDs imply that the material is very carbon-poor material \citep[e.g.,][]{2012MNRAS.424..333G,2012ApJ...750...69J}.
PG 1159 stars follow the [WC] central stars by no more than 10$^3$--10$^4$ yr (using a stellar evolutionary track for a 0.60-0.63~M$_\odot$ star from \citet{1993ApJ...413..641V} and known PG 1159 stellar parameters from \citet{1998A&A...334..618D}), while the oldest PG 1159 stars, those with no PNe, are another 10$^4$--10$^5$ yr older than that. Since the dust around Wolf-Rayet central stars may be distributed in long lived, Keplerian disks, it is possible that these disks have survived the hottest phases of the star and are present in the PG 1159 stage.

We have therefore obtained \emph{Spitzer}/IRS spectra of a sample of nine PG 1159 stars, all associated with PNe to look for evidence of dusty disks, increase the statistics of disks around CSPN, compare them further with debris disks around old WDs as well as with that around the Helix CSPN and study the disk longevity around CSPN.

\section{Observations and Reduction}

Nine PG 1159 stars, listed in Table 1,  were observed with the {\it Spitzer}/Infrared Spectrograph (IRS;  \citealt{Houck:2004lr}).  Staring-mode observations took place during IRS campaigns 58.1-58.3, 59.2 and 61.1 during 2008 December 5 to 9, 2009 January 9--25, March 5--8, and April 29.  All objects were observed with the Short-Low (SL; $5.2-14$ \micron; $\lambda/\Delta\lambda\sim90$) and Long-Low (LL; $14-38$ \micron; $\lambda/\Delta\lambda\sim90$) low-resolution modules.
The LL slit is 168\arcsec~x 10\farcs7 and the SL slit is 57\arcsec~x 3\farcs7. 

Our point-source extraction and calibration method has been extensively detailed previously \citep{Furlan:2006fj,Sargent:2006kx,Watson:2007qy,Sargent:2009uq}.  We start with the  \emph{Spitzer} Science Center IRS pipeline {\it Basic Calibrated Data} product (BCD) for each object.  The BCD is flat-fielded, dark-current subtracted, and stray-light corrected.  We employ a point-source spectral extraction and calibration method using the Spectral Modeling, Analysis and Reduction Tool (SMART; \citealt{Higdon:2004lr}) and IDL routines for post-pipeline processing.  In particular, we identify and correct for rogue (NaN) pixels in our two dimensional spectral data by linearly interpolating the four nearest neighboring pixel values.  To correct for the sky background, we subtract the off order spectrum 
($\sim$1--3\arcmin~away from target) in the same nod position of the on-target order.  The low-resolution, sky-subtracted spectra are then extracted using a variable-width extraction window that fits tightly to the IRS point-spread function.

To calibrate our spectra, we employ custom relative spectral response functions (RSRFs) which yield flux densities based on the signal detected at a given wavelength.  To produce the RSRFs, we use SMART to divide each of our spectra, nod by nod, for each order of each module.  A spectral template of a calibration star ($\alpha$ Lac; \citealt{Cohen:2003fk}) was identically prepared such that the quotient of the template and the observed stellar spectrum is the RSRF.  The low-resolution, sky-subtracted extractions were then multiplied by the RSRF corresponding to the relevant nod, order, and module.  In general, we found good agreement between flux values in regions of wavelength overlap between orders in each module.  For each object, the procedure described above produces a calibrated point-source spectrum. 

There are IRS/Peakup images (PUI; 16 \& 22 \micron) for the whole sample. In addition, three stars have IRAC (3.6, 4.5, 5.8, 8.0 \micron) images, and six have MIPS (24 \micron) images \citep[e.g.,][]{2011AJ....142...75C}. See Table 1 for a summary. These images have been examined for point sources at the coordinates of the sample stars. 
As an example, Figure 1 shows existing IR images of NGC 246.
NGC 246 and NGC 650 show a point source at the position of the star in all 4 IRAC bands.
The images also clearly show the nearby companions of these two stars. 
 JnEr 1 shows a point source only at 3.6 and 4.5 \micron. 
None of the stars with MIPS 24 \micron~imaging shows a point source at the location of the star. 
Also, there are no detections of point sources in any of the PUI images for the nine stars in the sample.

The IRS spectra for the sample stars are plotted in Figure 2. 
These spectra show strong emission lines due to the surrounding PNe \citep{2007ApJ...660.1282G}. The primary lines are [S IV] $\lambda$10.51, [Ne V] $\lambda$14.32, [Ne III] $\lambda$15.55, [S III] $\lambda$18.71, [Ne V] $\lambda$24.30, [O IV] $\lambda$25.89, and [S III] $\lambda$33.47 \micron. 
A point source extraction was attempted for both slits for each star, which are shown in Figure 2. The sky subtraction is taken $\sim$1\arcmin-3\arcmin~away from the star, well outside the slit position and outside the visible PN nebulosity. So the 1D spectra, extracted and shown in Figure 2, include any diffuse dust emission from the PN present in the portions of the slit that were extracted. The 2D spectra have point-like sources in the slit that have been extracted, but in each case these turned out to be dust condensations in the PN. Local sky subtraction was also attempted but no significant point sources centered on the target stars were detected.
As seen in Figure 2,  all of the stars in the sample show mid-IR emission rising toward longer wavelengths indicating the presence of cool dust.
The key question is, where is the dust, whose emission is being detected. Is the IR emission from a point source centered on the star in which case, it would be consistent with dust emission from a circumstellar disk, or is it diffuse emission from dust associated with the PN that surrounds each of these stars? 

Spectral energy distributions (SEDs) were constructed for the three stars, NGC 246, NGC 650, and JnEr 1,  with both IRAC and MIPS 24 \micron~photometry. 
SEDs for each of the three stars are shown in Figure 3, where the fluxes were all normalized to the distance of NGC 246 \citep[500 pc,][]{1999PASP..111..217B}. For the IRAC images, which overlap in wavelength with the IRS spectra, the brightness of the star is plotted both with sky subtraction (in an annulus close to the star) and with no sky subtraction. 
Since none of the stars showed a point source at 24 \micron, only the no-sky-subtraction 24 \micron~fluxes are plotted in Figure 3. 
The no-sky-subtraction fluxes closely mimic the fluxes of the IRS extractions. This clearly shows two important results. First, the sky-subtracted fluxes, in all three cases, are consistent with the tail of the blackbody SED from the star.  And second, the no-sky-subtraction fluxes agree very well with the IRS fluxes. So, the conclusion must be that the mid-IR emission from dust seen in the IRS spectra comes from diffuse emission in the PN surrounding each of the PG 1159 stars. 
This is reinforced by the fact that the six stars in our sample, that were also imaged with MIPS at 24 \micron, show no evidence for a point source. 
So we see no evidence for dusty disks in this small sample of PG 1159 stars.

In total, including our sample, twenty PG 1159 stars (ten with PNe and ten without) from the catalog of C. S. Jeffery\footnote{from http://www.arm.ac.uk/$\sim$csj/research/pg1159catalog.html} have been observed with \emph{Spitzer}/MIPS at 24 \micron. None of the PG 1159 stars with no PN, including PG 1159-035 itself, shows evidence for a dusty disk \citep{2011AJ....142...75C}. The ten stars  with PNe are listed in Table 1.

There is 2MASS $JHK$ photometry available for NGC 246 and NGC 650. $UBVRI$ photometry was available for NGC 246 \citep{1998MNRAS.294...93K}, and SDSS $ugriz$ photometry is available for NGC 650 and JnEr 1 \citep{2009ApJS..182..543A}. Low resolution {\it IUE} spectra were downloaded from the MAST archive for NGC 246. 
Photometry of NGC 246 and NGC 650 is difficult because each has close companions. NGC 246 has a physical G8-K0 V companion 3\farcs8 away \citep{1999PASP..111..217B}, and NGC 650 has two stars, thought to be background objects, only 1\farcs3 away \citep{1998A&A...335..277K}.




\section{The PG 1159 CSPN in the context of other CSPN}

The [WC] central stars of PN are thought to evolve to be PG 1159 stars with a PN,  then PG 1159 stars without a PN, and finally non-DA WDs \citep{2006PASP..118..183W}.  Dusty disks may be common around [WC] central stars, because they often exhibit dual-dust chemistry  \citep[the simultaneous presence of the carbon-rich PAHs and oxygen rich silicate dust; e.g.,][]{2002PASP..114..602D}, a phenomenon that has been attributed to the presence of an oxygen-rich disk coexisting alongside a carbon-rich outflow (e.g., Cohen et al. 1999). The fraction of disks among the [WC] central stars may be difficult to find using an infrared-excess technique, due to the particularly red colors of the hydrogen-deficient atmospheres of these stars. However, if present, they should be readily detected around the progeny of the [WC] stars, the PG 1159 stars.

For CSPN, in general, 18\% of 84 objects were detected to have a 24 \micron~flux excess \citep{2012ApJS..200....3B}. 
Of the ten PG 1159 stars with PNe observed at 24 \micron, five have such low S/N that they don't constrain the presence of a disk.  
One of the remaining five, the central star of the PN Abell 21, has been detected at 24 \micron~\citep{2011AJ....142...75C}. 
So, for this small sample, the detection rate of one in five objects is fully in line with the statistics of CSPN.  The fact that we did not detect a preponderance of circumstellar disks among the PG 1159 stars argues either in favor of their disks having been destroyed by intense radiation, that not all [WC] stars and, therefore, their descendant PG 1159 stars, have such disks, or that we are just observing the effect of low number statistics. 


\section{The classification of stars with disks}
\label{sec:mimic}

Before we discuss whether disks have a debris origin or not, we should clarify how the host stars should be classified. We have seen 24~\micron~excesses in WDs, both cool (old) and hot (young), and in CSPN. CSPN have very recently departed from the AGB and are at a similar evolutionary stage as the warm WDs. The cool WDs, on the other hand, left the AGB a long time ago.  

There is a second source of confusion. Some objects classified as PN were later found to be mimics \citep{2008PhDT.......109F,2010PASA...27..129F}. There are two types of mimics: young, hot WDs that lost their PN sometime ago, but are still hot and luminous enough to ionize the  interstellar medium, and stars that, by their temperature and gravity, can be placed on post-Red Giant Branch (RGB) evolutionary tracks. If hot enough, these stars can also ionize the interstellar medium and have a nebula that can be mistaken for a PN. An example of the former class is the disk object, EGB~1. Its nebulosity is not a bona fide PN, but is ionized interstellar medium around a young, hot WD; the central star of EGB~1 likely was surrounded by a PN until recently. Another similar mimic is the nebulosity discovered around the hot WD 0109+111 by \citet{1997A&A...327..721W}. Both EGB~1 and WD 0109+111 should be analyzed in the context of CSPN when it comes to  interpreting their disks.

The  latter class of mimics constitutes a third category of objects. An example is the disk object DeHt~5, which is traditionally listed as a PN \citep{2008PhDT.......109F,2013MNRAS.428.2118D}. The central star's temperature and gravity place this star very close to the post-RGB track of a $\sim$0.44~\msol\ star \citep{1999A&A...350..101N}, although \citet{2010ApJ...720..581G} find a mass of $\sim$0.54~\msol, more consistent with a post-AGB origin. As a mimic, this object is similar to EGB~5 and PHL~932 \citep[][neither of which is known to have a disk]{2011A&A...528L..16G}. \citet{2013MNRAS.428.2118D} detected a 4-$\sigma$ $J$ band excess for the central star of DeHt~5, consistent with an M5V companion. \citet{2012ApJS..200....3B} detected point sources in 2MASS, IRAC, and MIPS bands with a clear excess at 8 and 24 \micron. They concluded that there is a cool disk with a temperature of 190~K and a radius of 8.2~AU. Whatever the origin of this disk (most likely a stellar interaction that happened while the star was on the RGB, possibly a common envelope if the companion is found to orbit the star closely). 
\citet{2003MNRAS.341..477B} examined {\it HST}/STIS UV spectra and found hot, circumstellar gas, with lines redshifted relative to the WD rest frame, and speculate that this might indicate infalling matter.
This object is a third category of star with a dusty circumstellar environment alongside bona fide old WDs and younger, warmer WDs with or without a PN. 

Finally, we introduce the post-AGB stars. These are F or G supergiants that have departed the AGB, but are still much cooler than CSPN. These stars can have a pre-PN (a nebula that shines because of reflected light), in which case they are called central stars of pre-PN, or not, in which case we call them naked post-AGB stars \citep{2012IAUS..283..115B}.  


\section{Have any of the CSPN disks a debris nature?}

The last question we want to address is whether any of the disks found around CSPN are debris disks. While we acknowledge that CSPN debris disks cannot be like those around old, cool WDs, we wonder whether other CSPN disks could be formed by collisions of KBOs as was suggested for the Helix PN by Su et al. (2007). That scenario is plausible in the case of the Helix PN as investigated analytically by \citet{2010ApJ...715.1036D}.
Disks around central stars of PN can have formed from gas ejected during the AGB, plausibly, but not necessarily, because of a binary interaction, within which dust condensed. This is the likely origin of disks around naked post-AGB stars \citep{2009A&A...505.1221V}, central stars of pre-PN \citep[e.g.,][]{2012IAUS..283..115B} as well as disks around some CSPN \citep[e.g.,][]{2002ApJ...574L..83D}. 
Both scenarios have been amply discussed by \citet{2012ApJS..200....3B}, who point out the impossibility of discerning between them. Here, we examine the possibility that the Helix disk, as well as those detected around other PN central stars, have more plausibly an AGB origin than a shattered KBO one.

We present, in Figure~\ref{fig:CMD} a color-magnitude diagram of all the objects for which 8 and/or 24~\micron~imagery exists. All fluxes have been scaled to 500~pc using the distances listed in Table~2. The cool, old WDs have dim hot disks, as expected. The CSPN with data, on the other hand, reside in a strip extending from the top left to the bottom right of the figure, with the brightest members being much brighter, but only slightly redder than the cool WDs, and the dimmest members being only slightly brighter, but far redder than the cool WDs. The two brightest CSPN objects, NGC~2346 and NGC~6804, are known from independent observations to have large dusty tori: the former, a bright edge-on, bipolar PN, has photometric variability due to the close binary companion periodically hiding behind dust clumps in the disk \citep[][and references therein]{2001PASJ...53..901K}, while the latter has a possibly crystalline silicate feature in a Gemini/Michelle spectrum \citep{2012ApJS..200....3B}, indicating a long lived, Keplerian disk. The PG~1159 stars, A~21, and NGC~246 have only limits, and while the former has a  24~\micron~limit, which places it comfortably in the CSPN lane, the latter may be bluer than the strip. EGB~1 (which is a PN mimic, but whose central star is a hot and young WD, likely only slightly more evolved than a typical CSPN) is within the CSPN lane and finally, DeHt~5, which is also a PN mimic, but likely a post-RGB star,  also resides in the CSPN lane. The SED of CSPN disks are also compared to those of the debris disks around cool WDs in Figure~\ref{fig:SED}; the WD SEDs are flatter (hotter) and overall about 100 times dimmer. 

We also bring to bear the disks around post-AGB stars with no nebula \citep{2003ARA&A..41..391V}, and those with a pre-PN (\S~\ref{sec:mimic}), which are also plotted in Figure 4.  The plotted points for the naked post-AGB stars (with no nebula) are averages of approximately 10 objects with an oxygen-rich chemistry and 10 objects with a carbon-rich chemistry \citep{2011MNRAS.411.1597W}. The error bars indicate the range of the individual values. For the pre-PN group we use the NIR measurements from \citet[][2013, in preparation]{2012IAUS..283..115B}, extracted from interferometric observations that measured the small disk rather than the overall dusty pre-PN. All post-AGB disks, whether a pre-PN is present or not, cluster together. Post-AGB disks are brighter than all other disks. Their colors are intermediate between the hot WD debris disks and the cooler CSPN dim disks and similar to the colors of the brighter, hotter CSPN group. All post-AGB stars, whether with a pre-PN or not, are stars which are close to the AGB (temperatures of a few thousand Kelvin; Table 2). Their disks either just formed (as may be the case for the central stars of pre-PN) or even if they are older (as may be the case for the post-AGB with no nebula whose evolution may have been stalled by accretion; \citealt{2009A&A...505.1221V}), their Keplerian nature insures that they did not expand away \citep[][2013, in preparation]{2012IAUS..283..115B}.

We speculate here that the warm and bright disks around post-AGB stars, objects which have just left the AGB, become dim, red ones as the disk expands and disperses, and the star turns the ``knee" of the HR diagram and dims. 
Hotter central stars would have cooler disks if the hotter and dimmer star is destroying the nearby dust and the bulk of the emission is overall farther from the centre of radiation. 

We also notice that almost all the CSPN with 8 or 24 micron point source detections appear to be binaries or binary candidates. NGC~246, EGB~6, and K~1-22 are wide binaries (where EGB~6 has a known disk around the companion; \citealt{1999PASP..111..217B,2013ApJ...769...32L,1999AJ....118..488C}); 
NGC~2438 has a detected infrared excess consistent with an M3 V companion  \citep[although it could also be due to a warm disk -][]{2012ApJS..200....3B}. 
DeHt~5, EGB~1, and NGC~6853 have $J$-band flux excesses consistent with companions with spectral type M5 V \citep{2013MNRAS.428.2118D}, Sh~2-188 has a low-sigma detection of an $I$-band excess consistent with a companion of spectral type M4 V \citep{2013MNRAS.428.2118D}. NGC~2346 is a close, single-lined spectroscopic binary with an A5~V companion to a hot star \citep{1981ApJ...250..240M,1997Obs...117..338S}. 
The only objects in our sample that are not at present known to be binaries are NGC~6804, Abell~21, Sh 2-216, NGC 7139, and the Helix central star; the Helix central star has been thoroughly investigated and no companion has been detected down to a spectral type of mid-to-late M \citep{2007ApJ...657L..41S}. The lack of a detected companion does not prove the lack of a companion.  However, we cannot at this time state that all CSPN with disks are binaries. Detecting fainter and fainter companions as well as more disks will allow us to corroborate or dismiss an association between binarity and disks.
It is interesting to note that no cool WD with a dusty disks is known to have a close binary companion. One such WD binary \citep{2012ApJ...759...37D}, which was suggested to have a disk, seems instead to be magnetic \citep{2013MNRAS.436..241P}.



Finally, the low estimate of the dust mass of the Helix disk \citep{2007ApJ...657L..41S}, is about the same as that of the Solar System's Kuiper belt \citep[$\sim3\times10^{-7}$~M$_{\sun}$;][]{2007ApJ...657L..41S}, is 1000 times lower than the mass of the disk around the [WC10] CSPN CPD -56$\degree$ 8032 (Cohen et al. 1999; De Marco et al. 2002; $6\times10^{-4}$~M$_{\sun}$; an average of the two estimates of \citet{1999ApJ...513L.135C} and \citet{Clayton:2011lr}) and is also much less massive than the post-AGB disks for which we have data ($10^{-5} - 10^{-3}$~M$_{\sun}$; Bright 2013 in preparation). Disk dust masses determined for 11 disks around central stars of bona fide PN by Bilikova (PhD Thesis and in preparation) also show high values once all dust components are included.

\section{Conclusions}

Dusty disks have been detected around about 18\% of CSPN. One out of five PG 1159 stars with PNe \citep{2011AJ....142...75C,2012ApJS..200....3B} show evidence of a disk, in line with the rest of the PN group. This may indicate that PG~1159 stars do not show disks more often than non-PG 1159 CSPN. On the other hand, the statistics of PG1195 CSPN are extremely weak.
It is interesting that none of the PG1159 stars, which have no PN and are therefore 10$^4$--10$^5$ yr older, shows evidence for a dusty disk.

The characteristics of the disks, such as mass, radius, composition, and temperature will yield their origin. Currently, however, we have only partial information on disks around a range of object classes, which makes it difficult to draw accurate conclusions. While the rare dust disks found around old DZ WDs are certainly of a different nature from those around CSPN, it is still unclear what the nature of the CSPN disks is: 
do they derive from the pulverization of KBOs, as is plausible for the Helix PN, or are they dust formed in the ejecta of AGB stars? 
Based on the colors and brightness of the CSPN disks relative to those of stars that have just left the AGB, one may argue that the origin of CSPN disks is from AGB ejecta. 

One may also argue that CSPN disks may themselves have more than one origin. Disks around hydrogen deficient [WC] and PG 1159 CSPN could derive from a different evolution from those around hydrogen-normal CSPN and cooler post-AGB stars (with or without PN). On the other hand, on the color-magnitude diagram, the Helix CSPN disk resides close to other hot CSPN, possibly indicating a common origin.

Finally, it came as a surprise that 8 out of 13 CSPN with disks detected because of 8 or 24-\micron~excess are either binaries or likely/possible binaries. This connection, which should be further explored, may argue for an AGB origin.

\acknowledgments
This work was supported by \emph{Spitzer Space Telescope} RSA No.~1364990 issued by
Caltech/JPL. JN is supported by NSF AAP Fellowship AST-1102738 and by NASA HST grant AR-12146.04-A. OD gratefully acknowledges support from Australian Research Council Future Fellowship grant FT120100452.
TR is supported by the German Aerospace Center (DLR, grant 05\,OR\,1301).

\bibliography{/Users/gclayton/projects/latexstuff/everything2}

\begin{thebibliography}{70}
\expandafter\ifx\csname natexlab\endcsname\relax\def\natexlab#1{#1}\fi

\bibitem[{{Aannestad} {et~al.}(1993){Aannestad}, {Kenyon}, {Hammond}, \&
  {Sion}}]{1993AJ....105.1033A}
{Aannestad}, P.~A., {Kenyon}, S.~J., {Hammond}, G.~L., \& {Sion}, E.~M. 1993,
  \aj, 105, 1033

\bibitem[{{Abazajian} {et~al.}(2009){Abazajian}, {Adelman-McCarthy},
  {Ag{\"u}eros}, {Allam}, {Allende Prieto}, {An}, {Anderson}, {Anderson},
  {Annis}, {Bahcall}, \& et~al.}]{2009ApJS..182..543A}
{Abazajian}, K.~N., {et~al.} 2009, \apjs, 182, 543

\bibitem[{{Bannister} {et~al.}(2003){Bannister}, {Barstow}, {Holberg}, \&
  {Bruhweiler}}]{2003MNRAS.341..477B}
{Bannister}, N.~P., {Barstow}, M.~A., {Holberg}, J.~B., \& {Bruhweiler}, F.~C.
  2003, \mnras, 341, 477

\bibitem[{{Barber} {et~al.}(2012){Barber}, {Patterson}, {Kilic}, {Leggett},
  {Dufour}, {Bloom}, \& {Starr}}]{2012ApJ...760...26B}
{Barber}, S.~D., {Patterson}, A.~J., {Kilic}, M., {Leggett}, S.~K., {Dufour},
  P., {Bloom}, J.~S., \& {Starr}, D.~L. 2012, \apj, 760, 26

\bibitem[{{Bil{\'{\i}}kov{\'a}} {et~al.}(2012){Bil{\'{\i}}kov{\'a}}, {Chu},
  {Gruendl}, {Su}, \& {De Marco}}]{2012ApJS..200....3B}
{Bil{\'{\i}}kov{\'a}}, J., {Chu}, Y.-H., {Gruendl}, R.~A., {Su}, K.~Y.~L., \&
  {De Marco}, O. 2012, \apjs, 200, 3

\bibitem[{{Bond} \& {Ciardullo}(1999)}]{1999PASP..111..217B}
{Bond}, H.~E., \& {Ciardullo}, R. 1999, \pasp, 111, 217

\bibitem[{{Bright} {et~al.}(2012){Bright}, {De Marco}, {Chesneau}, {Lagadec},
  {Van Winckel}, \& {Hrivnak}}]{2012IAUS..283..115B}
{Bright}, S.~N., {De Marco}, O., {Chesneau}, O., {Lagadec}, E., {Van Winckel},
  H., \& {Hrivnak}, B.~J. 2012, in IAU Symposium, Vol. 283, 115

\bibitem[{{Brinkworth} {et~al.}(2012){Brinkworth}, {G{\"a}nsicke}, {Girven},
  {Hoard}, {Marsh}, {Parsons}, \& {Koester}}]{2012ApJ...750...86B}
{Brinkworth}, C.~S., {G{\"a}nsicke}, B.~T., {Girven}, J.~M., {Hoard}, D.~W.,
  {Marsh}, T.~R., {Parsons}, S.~G., \& {Koester}, D. 2012, \apj, 750, 86

\bibitem[{{Brinkworth} {et~al.}(2009){Brinkworth}, {G{\"a}nsicke}, {Marsh},
  {Hoard}, \& {Tappert}}]{2009ApJ...696.1402B}
{Brinkworth}, C.~S., {G{\"a}nsicke}, B.~T., {Marsh}, T.~R., {Hoard}, D.~W., \&
  {Tappert}, C. 2009, \apj, 696, 1402

\bibitem[{{Chu} {et~al.}(2011){Chu}, {Su}, {Bilikova}, {Gruendl}, {De Marco},
  {Guerrero}, {Updike}, {Volk}, \& {Rauch}}]{2011AJ....142...75C}
{Chu}, Y.-H., {et~al.} 2011, \aj, 142, 75

\bibitem[{{Ciardullo} {et~al.}(1999){Ciardullo}, {Bond}, {Sipior}, {Fullton},
  {Zhang}, \& {Schaefer}}]{1999AJ....118..488C}
{Ciardullo}, R., {Bond}, H.~E., {Sipior}, M.~S., {Fullton}, L.~K., {Zhang},
  C.-Y., \& {Schaefer}, K.~G. 1999, \aj, 118, 488

\bibitem[{{Clayton} {et~al.}(2011){Clayton}, {De Marco}, {Whitney}, {Babler},
  {Gallagher}, {Nordhaus}, {Speck}, {Wolff}, {Freeman}, {Camp}, {Lawson},
  {Roman-Duval}, {Misselt}, {Meade}, {Sonneborn}, {Matsuura}, \&
  {Meixner}}]{Clayton:2011lr}
{Clayton}, G.~C., {et~al.} 2011, \aj, 142, 54

\bibitem[{{Cohen} {et~al.}(1999){Cohen}, {Barlow}, {Sylvester}, {Liu}, {Cox},
  {Lim}, {Schmitt}, \& {Speck}}]{1999ApJ...513L.135C}
{Cohen}, M., {Barlow}, M.~J., {Sylvester}, R.~J., {Liu}, X.-W., {Cox}, P.,
  {Lim}, T., {Schmitt}, B., \& {Speck}, A.~K. 1999, \apjl, 513, L135

\bibitem[{{Cohen} {et~al.}(2003){Cohen}, {Megeath}, {Hammersley},
  {Mart{\'{\i}}n-Luis}, \& {Stauffer}}]{Cohen:2003fk}
{Cohen}, M., {Megeath}, S.~T., {Hammersley}, P.~L., {Mart{\'{\i}}n-Luis}, F.,
  \& {Stauffer}, J. 2003, \aj, 125, 2645

\bibitem[{{De Marco} {et~al.}(2002){De Marco}, {Barlow}, \&
  {Cohen}}]{2002ApJ...574L..83D}
{De Marco}, O., {Barlow}, M.~J., \& {Cohen}, M. 2002, \apjl, 574, L83

\bibitem[{{De Marco} {et~al.}(2013){De Marco}, {Passy}, {Frew}, {Moe}, \&
  {Jacoby}}]{2013MNRAS.428.2118D}
{De Marco}, O., {Passy}, J.-C., {Frew}, D.~J., {Moe}, M., \& {Jacoby}, G.~H.
  2013, \mnras, 428, 2118

\bibitem[{{De Marco} \& {Soker}(2002)}]{2002PASP..114..602D}
{De Marco}, O., \& {Soker}, N. 2002, \pasp, 114, 602

\bibitem[{{Debes} {et~al.}(2012){Debes}, {Hoard}, {Farihi}, {Wachter},
  {Leisawitz}, \& {Cohen}}]{2012ApJ...759...37D}
{Debes}, J.~H., {Hoard}, D.~W., {Farihi}, J., {Wachter}, S., {Leisawitz},
  D.~T., \& {Cohen}, M. 2012, \apj, 759, 37

\bibitem[{{Dong} {et~al.}(2010){Dong}, {Wang}, {Lin}, \&
  {Liu}}]{2010ApJ...715.1036D}
{Dong}, R., {Wang}, Y., {Lin}, D.~N.~C., \& {Liu}, X.-W. 2010, \apj, 715, 1036

\bibitem[{{Dreizler} \& {Heber}(1998)}]{1998A&A...334..618D}
{Dreizler}, S., \& {Heber}, U. 1998, \aap, 334, 618

\bibitem[{{Dupuis} {et~al.}(1993){Dupuis}, {Fontaine}, {Pelletier}, \&
  {Wesemael}}]{1993ApJS...84...73D}
{Dupuis}, J., {Fontaine}, G., {Pelletier}, C., \& {Wesemael}, F. 1993, \apjs,
  84, 73

\bibitem[{{Farihi} {et~al.}(2009){Farihi}, {Jura}, \&
  {Zuckerman}}]{2009ApJ...694..805F}
{Farihi}, J., {Jura}, M., \& {Zuckerman}, B. 2009, \apj, 694, 805

\bibitem[{{Frew}(2008)}]{2008PhDT.......109F}
{Frew}, D.~J. 2008, PhD thesis, Department of Physics, Macquarie University,
  NSW 2109, Australia

\bibitem[{{Frew} \& {Parker}(2010)}]{2010PASA...27..129F}
{Frew}, D.~J., \& {Parker}, Q.~A. 2010, PASA, 27, 129

\bibitem[{{Furlan} {et~al.}(2006){Furlan}, {Hartmann}, {Calvet}, {D'Alessio},
  {Franco-Hern{\'a}ndez}, {Forrest}, {Watson}, {Uchida}, {Sargent}, {Green},
  {Keller}, \& {Herter}}]{Furlan:2006fj}
{Furlan}, E., {et~al.} 2006, \apjs, 165, 568

\bibitem[{{G{\"a}nsicke} {et~al.}(2012){G{\"a}nsicke}, {Koester}, {Farihi},
  {Girven}, {Parsons}, \& {Breedt}}]{2012MNRAS.424..333G}
{G{\"a}nsicke}, B.~T., {Koester}, D., {Farihi}, J., {Girven}, J., {Parsons},
  S.~G., \& {Breedt}, E. 2012, \mnras, 424, 333

\bibitem[{{G{\"a}nsicke} {et~al.}(2006){G{\"a}nsicke}, {Marsh}, {Southworth},
  \& {Rebassa-Mansergas}}]{2006Sci...314.1908G}
{G{\"a}nsicke}, B.~T., {Marsh}, T.~R., {Southworth}, J., \&
  {Rebassa-Mansergas}, A. 2006, Science, 314, 1908

\bibitem[{{Geier} {et~al.}(2011){Geier}, {Napiwotzki}, {Heber}, \&
  {Nelemans}}]{2011A&A...528L..16G}
{Geier}, S., {Napiwotzki}, R., {Heber}, U., \& {Nelemans}, G. 2011, \aap, 528,
  L16

\bibitem[{{Gianninas} {et~al.}(2010){Gianninas}, {Bergeron}, {Dupuis}, \&
  {Ruiz}}]{2010ApJ...720..581G}
{Gianninas}, A., {Bergeron}, P., {Dupuis}, J., \& {Ruiz}, M.~T. 2010, \apj,
  720, 581

\bibitem[{{Girven} {et~al.}(2011){Girven}, {G{\"a}nsicke}, {Steeghs}, \&
  {Koester}}]{2011MNRAS.417.1210G}
{Girven}, J., {G{\"a}nsicke}, B.~T., {Steeghs}, D., \& {Koester}, D. 2011,
  \mnras, 417, 1210

\bibitem[{{Graham} {et~al.}(1990){Graham}, {Matthews}, {Neugebauer}, \&
  {Soifer}}]{1990ApJ...357..216G}
{Graham}, J.~R., {Matthews}, K., {Neugebauer}, G., \& {Soifer}, B.~T. 1990,
  \apj, 357, 216

\bibitem[{{Guiles} {et~al.}(2007){Guiles}, {Bernard-Salas}, {Pottasch}, \&
  {Roellig}}]{2007ApJ...660.1282G}
{Guiles}, S., {Bernard-Salas}, J., {Pottasch}, S.~R., \& {Roellig}, T.~L. 2007,
  \apj, 660, 1282

\bibitem[{{Higdon} {et~al.}(2004){Higdon}, {Devost}, {Higdon}, {Brandl},
  {Houck}, {Hall}, {Barry}, {Charmandaris}, {Smith}, {Sloan}, \&
  {Green}}]{Higdon:2004lr}
{Higdon}, S.~J.~U., {et~al.} 2004, \pasp, 116, 975

\bibitem[{{Hodge} {et~al.}(2004){Hodge}, {Kraemer}, {Price}, \&
  {Walker}}]{2004ApJS..151..299H}
{Hodge}, T.~M., {Kraemer}, K.~E., {Price}, S.~D., \& {Walker}, H.~J. 2004,
  \apjs, 151, 299

\bibitem[{{Houck} {et~al.}(2004){Houck}, {Roellig}, {van Cleve}, {Forrest},
  {Herter}, {Lawrence}, {Matthews}, {Reitsema}, {Soifer}, {Watson}, {Weedman},
  {Huisjen}, {Troeltzsch}, {Barry}, {Bernard-Salas}, {Blacken}, {Brandl},
  {Charmandaris}, {Devost}, {Gull}, {Hall}, {Henderson}, {Higdon}, {Pirger},
  {Schoenwald}, {Sloan}, {Uchida}, {Appleton}, {Armus}, {Burgdorf},
  {Fajardo-Acosta}, {Grillmair}, {Ingalls}, {Morris}, \&
  {Teplitz}}]{Houck:2004lr}
{Houck}, J.~R., {et~al.} 2004, \apjs, 154, 18

\bibitem[{{Hu} {et~al.}(1993){Hu}, {Slijkhuis}, {de Jong}, \&
  {Jiang}}]{1993A&AS..100..413H}
{Hu}, J.~Y., {Slijkhuis}, S., {de Jong}, T., \& {Jiang}, B.~W. 1993, \aaps,
  100, 413

\bibitem[{{Jura}(2003)}]{2003ApJ...584L..91J}
{Jura}, M. 2003, \apjl, 584, L91

\bibitem[{{Jura} {et~al.}(2007{\natexlab{a}}){Jura}, {Farihi}, \&
  {Zuckerman}}]{2007ApJ...663.1285J}
{Jura}, M., {Farihi}, J., \& {Zuckerman}, B. 2007{\natexlab{a}}, \apj, 663,
  1285

\bibitem[{{Jura} {et~al.}(2007{\natexlab{b}}){Jura}, {Farihi}, {Zuckerman}, \&
  {Becklin}}]{2007AJ....133.1927J}
{Jura}, M., {Farihi}, J., {Zuckerman}, B., \& {Becklin}, E.~E.
  2007{\natexlab{b}}, \aj, 133, 1927

\bibitem[{{Jura} {et~al.}(2012){Jura}, {Xu}, {Klein}, {Koester}, \&
  {Zuckerman}}]{2012ApJ...750...69J}
{Jura}, M., {Xu}, S., {Klein}, B., {Koester}, D., \& {Zuckerman}, B. 2012,
  \apj, 750, 69

\bibitem[{{Kato} {et~al.}(2001){Kato}, {Nogami}, \&
  {Baba}}]{2001PASJ...53..901K}
{Kato}, T., {Nogami}, D., \& {Baba}, H. 2001, \pasj, 53, 901

\bibitem[{{Kilkenny} {et~al.}(1998){Kilkenny}, {van Wyk}, {Roberts}, {Marang},
  \& {Cooper}}]{1998MNRAS.294...93K}
{Kilkenny}, D., {van Wyk}, F., {Roberts}, G., {Marang}, F., \& {Cooper}, D.
  1998, \mnras, 294, 93

\bibitem[{{Koesterke}(2001)}]{2001Ap&SS.275...41K}
{Koesterke}, L. 2001, \apss, 275, 41

\bibitem[{{Koornneef} \& {Pottasch}(1998)}]{1998A&A...335..277K}
{Koornneef}, J., \& {Pottasch}, S.~R. 1998, \aap, 335, 277

\bibitem[{{Liebert} {et~al.}(2013){Liebert}, {Bond}, {Dufour}, {Ciardullo},
  {Meakes}, {Renzini}, \& {Gianninas}}]{2013ApJ...769...32L}
{Liebert}, J., {Bond}, H.~E., {Dufour}, P., {Ciardullo}, R., {Meakes}, M.~G.,
  {Renzini}, A., \& {Gianninas}, A. 2013, \apj, 769, 32

\bibitem[{{Matsuura} {et~al.}(2004){Matsuura}, {Zijlstra}, {Molster}, {Hony},
  {Waters}, {Kemper}, {Bowey}, {Chihara}, {Koike}, \&
  {Keller}}]{2004ApJ...604..791M}
{Matsuura}, M., {et~al.} 2004, \apj, 604, 791

\bibitem[{{Mendez} \& {Niemela}(1981)}]{1981ApJ...250..240M}
{Mendez}, R.~H., \& {Niemela}, V.~S. 1981, \apj, 250, 240

\bibitem[{{Napiwotzki}(1999)}]{1999A&A...350..101N}
{Napiwotzki}, R. 1999, \aap, 350, 101

\bibitem[{{Parsons} {et~al.}(2013){Parsons}, {Marsh}, {G{\"a}nsicke},
  {Schreiber}, {Bours}, {Dhillon}, \& {Littlefair}}]{2013MNRAS.436..241P}
{Parsons}, S.~G., {Marsh}, T.~R., {G{\"a}nsicke}, B.~T., {Schreiber}, M.~R.,
  {Bours}, M.~C.~P., {Dhillon}, V.~S., \& {Littlefair}, S.~P. 2013, \mnras,
  436, 241

\bibitem[{{Perea-Calder{\'o}n} {et~al.}(2009){Perea-Calder{\'o}n},
  {Garc{\'{\i}}a-Hern{\'a}ndez}, {Garc{\'{\i}}a-Lario}, {Szczerba}, \&
  {Bobrowsky}}]{2009A&A...495L...5P}
{Perea-Calder{\'o}n}, J.~V., {Garc{\'{\i}}a-Hern{\'a}ndez}, D.~A.,
  {Garc{\'{\i}}a-Lario}, P., {Szczerba}, R., \& {Bobrowsky}, M. 2009, \aap,
  495, L5

\bibitem[{{Phillips}(2003)}]{2003MNRAS.344..501P}
{Phillips}, J.~P. 2003, \mnras, 344, 501

\bibitem[{{Reach} {et~al.}(2005){Reach}, {Kuchner}, {von Hippel}, {Burrows},
  {Mullally}, {Kilic}, \& {Winget}}]{2005ApJ...635L.161R}
{Reach}, W.~T., {Kuchner}, M.~J., {von Hippel}, T., {Burrows}, A., {Mullally},
  F., {Kilic}, M., \& {Winget}, D.~E. 2005, \apjl, 635, L161

\bibitem[{{Sargent} {et~al.}(2006){Sargent}, {Forrest}, {D'Alessio}, {Li},
  {Najita}, {Watson}, {Calvet}, {Furlan}, {Green}, {Kim}, {Sloan}, {Chen},
  {Hartmann}, \& {Houck}}]{Sargent:2006kx}
{Sargent}, B., {et~al.} 2006, \apj, 645, 395

\bibitem[{{Sargent} {et~al.}(2009){Sargent}, {Forrest}, {Tayrien}, {McClure},
  {Li}, {Basu}, {Manoj}, {Watson}, {Bohac}, {Furlan}, {Kim}, {Green}, \&
  {Sloan}}]{Sargent:2009uq}
{Sargent}, B.~A., {et~al.} 2009, \apj, 690, 1193

\bibitem[{{Slijkhuis} {et~al.}(1991){Slijkhuis}, {de Jong}, \&
  {Hu}}]{1991A&A...248..547S}
{Slijkhuis}, S., {de Jong}, T., \& {Hu}, J.~Y. 1991, \aap, 248, 547

\bibitem[{{Smalley}(1997)}]{1997Obs...117..338S}
{Smalley}, B. 1997, The Observatory, 117, 338

\bibitem[{{Steele} {et~al.}(2011){Steele}, {Burleigh}, {Dobbie}, {Jameson},
  {Barstow}, \& {Satterthwaite}}]{2011MNRAS.416.2768S}
{Steele}, P.~R., {Burleigh}, M.~R., {Dobbie}, P.~D., {Jameson}, R.~F.,
  {Barstow}, M.~A., \& {Satterthwaite}, R.~P. 2011, \mnras, 416, 2768

\bibitem[{{Su} {et~al.}(2007){Su}, {Chu}, {Rieke}, {Huggins}, {Gruendl},
  {Napiwotzki}, {Rauch}, {Latter}, \& {Volk}}]{2007ApJ...657L..41S}
{Su}, K.~Y.~L., {et~al.} 2007, \apjl, 657, L41

\bibitem[{{Tweedy} \& {Kwitter}(1994)}]{1994ApJ...433L..93T}
{Tweedy}, R.~W., \& {Kwitter}, K.~B. 1994, \apjl, 433, L93

\bibitem[{{van Winckel}(2003)}]{2003ARA&A..41..391V}
{van Winckel}, H. 2003, \araa, 41, 391

\bibitem[{{van Winckel} {et~al.}(2009){van Winckel}, {Lloyd Evans}, {Briquet},
  {De Cat}, {Degroote}, {De Meester}, {De Ridder}, {Deroo}, {Desmet},
  {Drummond}, {Eyer}, {Groenewegen}, {Kolenberg}, {Kilkenny}, {Ladjal},
  {Lefever}, {Maas}, {Marang}, {Martinez}, {{\O}stensen}, {Raskin}, {Reyniers},
  {Royer}, {Saesen}, {Uytterhoeven}, {Vanautgaerden}, {Vandenbussche}, {van
  Wyk}, {Vu{\v c}kovi{\'c}}, {Waelkens}, \& {Zima}}]{2009A&A...505.1221V}
{van Winckel}, H., {et~al.} 2009, \aap, 505, 1221

\bibitem[{{Vassiliadis} \& {Wood}(1993)}]{1993ApJ...413..641V}
{Vassiliadis}, E., \& {Wood}, P.~R. 1993, \apj, 413, 641

\bibitem[{{Watson} {et~al.}(2007){Watson}, {Bohac}, {Hull}, {Forrest},
  {Furlan}, {Najita}, {Calvet}, {D'Alessio}, {Hartmann}, {Sargent}, {Green},
  {Kim}, \& {Houck}}]{Watson:2007qy}
{Watson}, D.~M., {et~al.} 2007, \nat, 448, 1026

\bibitem[{{Weidemann}(1960)}]{1960ApJ...131..638W}
{Weidemann}, V. 1960, \apj, 131, 638

\bibitem[{{Werner} {et~al.}(1997){Werner}, {Bagschik}, {Rauch}, \&
  {Napiwotzki}}]{1997A&A...327..721W}
{Werner}, K., {Bagschik}, K., {Rauch}, T., \& {Napiwotzki}, R. 1997, \aap, 327,
  721

\bibitem[{{Werner} \& {Herwig}(2006)}]{2006PASP..118..183W}
{Werner}, K., \& {Herwig}, F. 2006, \pasp, 118, 183

\bibitem[{{Werner} {et~al.}(2004){Werner}, {Rauch}, {Reiff}, {Kruk}, \&
  {Napiwotzki}}]{2004A&A...427..685W}
{Werner}, K., {Rauch}, T., {Reiff}, E., {Kruk}, J.~W., \& {Napiwotzki}, R.
  2004, \aap, 427, 685

\bibitem[{{Woods} {et~al.}(2011){Woods}, {Oliveira}, {Kemper}, {van Loon},
  {Sargent}, {Matsuura}, {Szczerba}, {Volk}, {Zijlstra}, {Sloan}, {Lagadec},
  {McDonald}, {Jones}, {Gorjian}, {Kraemer}, {Gielen}, {Meixner}, {Blum},
  {Sewi{\l}o}, {Riebel}, {Shiao}, {Chen}, {Boyer}, {Indebetouw}, {Antoniou},
  {Bernard}, {Cohen}, {Dijkstra}, {Galametz}, {Galliano}, {Gordon}, {Harris},
  {Hony}, {Hora}, {Kawamura}, {Lawton}, {Leisenring}, {Madden}, {Marengo},
  {McGuire}, {Mulia}, {O'Halloran}, {Olsen}, {Paladini}, {Paradis}, {Reach},
  {Rubin}, {Sandstrom}, {Soszy{\'n}ski}, {Speck}, {Srinivasan}, {Tielens}, {van
  Aarle}, {van Dyk}, {van Winckel}, {Vijh}, {Whitney}, \&
  {Wilkins}}]{2011MNRAS.411.1597W}
{Woods}, P.~M., {et~al.} 2011, \mnras, 411, 1597

\bibitem[{{Xu} \& {Jura}(2012)}]{2012ApJ...745...88X}
{Xu}, S., \& {Jura}, M. 2012, \apj, 745, 88

\bibitem[{{Zuckerman} \& {Becklin}(1987)}]{1987Natur.330..138Z}
{Zuckerman}, B., \& {Becklin}, E.~E. 1987, \nat, 330, 138

\end{thebibliography}


\begin{deluxetable}{llcccc}
\tablecaption{PG 1159 Stars with PNe observed with Spitzer$^a$}
\tablenum{1}
\tablehead{\colhead{PN name}&\colhead{Star name}&\colhead{T$_{eff}$ (kK)}&
           \colhead{d (pc)}&
           \colhead{IRAC$^b$}&
            \colhead{MIPS/24$^c$ (mJy)} }
\startdata
NGC 246$^{d,e}$&WD~0044-121&150&495&yes&$<$26.2\\
NGC 650$^{d}$&WD~0139+513&140&746&yes&$<$30\\
StWr~3-2&Wray 17-1$^{d}$&140&\nodata&\nodata&\nodata\\
Abell 21$^{e}$&WD~0726+133&140&541&yes&0.916 $\pm$ 0.114\\
JnEr 1$^{d,e}$&WD~0753+535&130&1145&yes&$<$0.675\\
Lo 4$^{d,e}$&WD~1003-441&120&3000&\nodata&$<$19.9\\
Abell 43$^{d,e}$&WD~1751+106&110&2050&\nodata&$<$6.1\\
NGC 6765$^{d}$&WD~1909+304&\nodata&2334&\nodata&\nodata\\
NGC 6852$^{d,e}$&WD~1958+015&\nodata&2736&\nodata&$<$140.4\\
NGC 7094$^{d}$&WD~2134+125&110&1390&\nodata&\nodata\\
Jn 1$^{e}$&WD~2333+301&170&&\nodata&$<$0.824\\
Jacoby~1 & WD 1520+525$^{e}$&150&600&\nodata&$<$0.358\\
MWP 1$^{e}$&WD~2115+339&170&1400&\nodata&$<$0.903
\enddata

\tablenotetext{a}{from http://www.arm.ac.uk/$\sim$csj/research/pg1159catalog.html}
\tablenotetext{b}{IRAC imaging is available.}
\tablenotetext{c}{MIPS 24 \micron~flux estimates are from  \citet{2011AJ....142...75C} except for NGC 650 which is from this study.}
\tablenotetext{d}{This study}  
\tablenotetext{e}{\citet{2011AJ....142...75C}, \citet{2012ApJS..200....3B}}

\end{deluxetable}


\begin{deluxetable}{llccccll}
\tablecaption{Stars for [8] vs. [8-24] Plot}
\tablenum{2}
\tablehead{\colhead{Star name}&\colhead{Nebula Name}&\colhead{Star Type}&
           \colhead{T$_{\rm eff}$}&
           \colhead{d}&
           \colhead{[8]$^a$}&
            \colhead{[8-24]$^a$} &
            \colhead{Ref.$^b$}\\
             \colhead{}& \colhead{}&\colhead{}&\colhead{(kK)}&\colhead{(pc)}&\colhead{(mag)}&\colhead{(mag)}&\colhead{}
            }
\startdata
GD 133     &--&  old WD  &12.2&43     &  18.18    &   1.95&2, C\\
GD 56     &--&  old WD  &14.4&74    &   16.05   &   0.72&2, C\\
WD~2326+049&--& old WD&11.8&14&17.50&1.05&7, K\\
WD~0146+187&--& old WD&11.5&48&17.98&0.95&8, L\\
WD~2115-560&--& old WD&9.7&22&18.94&0.93&8, L\\
WD~1729+371&--& old WD&10.5&57&17.21&1.22&9, L\\
WD~0726+133&Abell 21   &  CSPN, PG 1159&140& 540 &   $ >$15.64   &    $>$6.08&1, A\\
WD~0044-121&NGC 246&CSPN, PG 1159$^c$&150&500&12.74&$<$6.65&1, B\\
&Helix  &CSPN$^d$  &110&    215      & 15.74    &   8.49&1, 3, D\\
WD~1957+225&NGC 6853    &   CSPN$^e$  &130&379    &   12.95   &    6.12&1, E\\
HD~293373&NGC 2346    &   CSPN$^f$  &100&900     &  4.36   &    1.58&1, F\\
HD~183932&NGC 6804     &  CSPN$^g$ &85& 1470  &     4.51    &   2.79&1, G\\
WD~0103+732&EGB 1 &  young WD; CSPN mimic$^e$ &147&650&13.65&5.68&1, J\\
WD~0127+581&Sh 2-188 &  CSPN$^e$ &102&600&14.21&3.80&1, J\\
WD~0439+466&Sh 2-216&CSPN&95&129&16.82&6.65&1, J\\
&K 1-22 &  CSPN$^c$ &141&1330&10.12&2.69&1, J\\
WD~0950+139&EGB 6 &  CSPN$^c$ &108&645&10.02&3.62&1, J\\
&NGC 7139&CSPN&117&2400&10.27&\nodata&1, D\\
&NGC 2438 &  CSPN$^h$ &114&1200&12.54&7.54&1, J\\
&IRAS 16279-4757$^i$  &   Pre-PN  &5.7&2000 &-1.87 &4.48&4, H\\
&IRAS 08005-2356$^i$    & Pre-PN$^j$ &6.8&4000 & -2.84&3.13&5, I\\
&Average O-rich&Post-AGB&\nodata&50,000 &-1.50 &2.50&6\\
&Average C-rich&Post-AGB&\nodata&50,000 & -2.50&4.50&6\\
&Average C-rich&Post-AGB&\nodata&50,000 & -2.50&4.50&6\\
&DeHt~5$^k$    &   Post-RGB$^e$ &70 &345  &     15.51    &   5.56&1, D
\enddata
\tablenotetext{a}{All fluxes have been scaled to 500 pc and converted to magnitudes.}
\tablenotetext{b}{Flux references: 1) \citet{2012ApJS..200....3B}, 2) \citet{2007ApJ...663.1285J}, 3) \citet{2007ApJ...657L..41S}, 4) \citet{2004ApJ...604..791M}, 5) \citet{1991A&A...248..547S}, 6) \citet{2011MNRAS.411.1597W}, 7)  \citet{2005ApJ...635L.161R}, 8) \citet{2009ApJ...694..805F}, 9) \citet{2007AJ....133.1927J}.\\
T$_{eff}$  references: A) \citet{2004A&A...427..685W}, B) \citet{2001Ap&SS.275...41K}, C) \citet{2007ApJ...663.1285J}, 
D) \citet{2008PhDT.......109F}, E) \citet{1999A&A...350..101N}, F) \citet{1981ApJ...250..240M}, G) \citet{2003MNRAS.344..501P}, H)  \citet{1993A&AS..100..413H}, I) \citet{1991A&A...248..547S}, J) \citet{2012ApJS..200....3B}, K) \citet{2005ApJ...635L.161R}, L) \citet{2009ApJ...694..805F}.}
\tablenotetext{c}{Wide binaries \citealt{1999AJ....118..488C,2013ApJ...769...32L}.}
\tablenotetext{d}{No companion brighter than mid-to-late M \citep{2007ApJ...657L..41S}.}
\tablenotetext{e}{$I$- and $J$-band excess binary candidates \citep{2013MNRAS.428.2118D}.}
\tablenotetext{f}{Close binary \citep{1981ApJ...250..240M}.}
\tablenotetext{g}{Suspected binary \citep{2008PhDT.......109F}.}
\tablenotetext{h}{Infrared excess binary candidate \citep{2012ApJS..200....3B}.}
 \tablenotetext{i}{Used MSX A and E fluxes instead of IRAC 8 and 24 \micron.}
 \tablenotetext{j}{Suspected binary (Bright 2013, in preparation).}
\tablenotetext{k}{PN mimic; see Section 4.2.}
\end{deluxetable}

\begin{figure}
\figurenum{1} 
\begin{center}
\includegraphics[width=6in,angle=0]{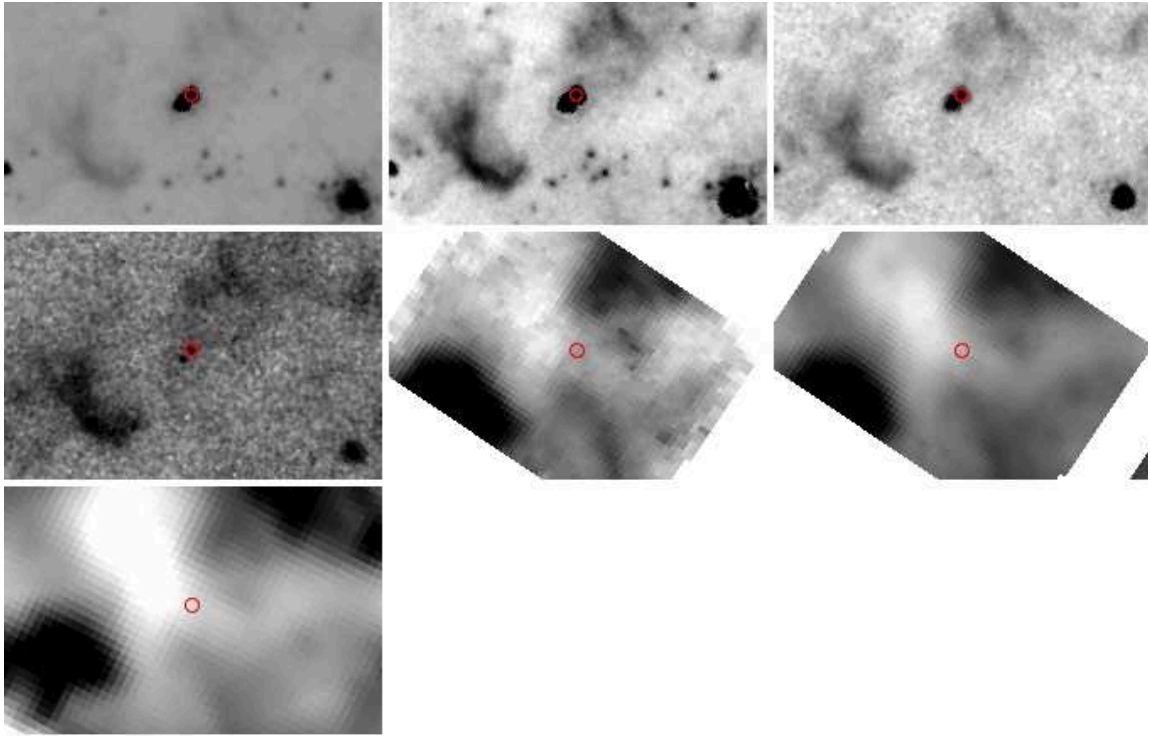}\\
\end{center}
\caption{NGC 246. IRAC 3.6, 4.5, 5.8, 8.0 \micron\ (upper panels and middle-left panel), PUI 16 \& 22 \micron\ (rightmost middle panels), and MIPS 24 \micron\ (lower panel) images. The red circle marks the position of the central star.  North is up and East is to the left. The long axis of each field is 1\farcm85.}
\end{figure}

\begin{figure}
\figurenum{2} 
\begin{center}
\includegraphics[width=7in,angle=0]{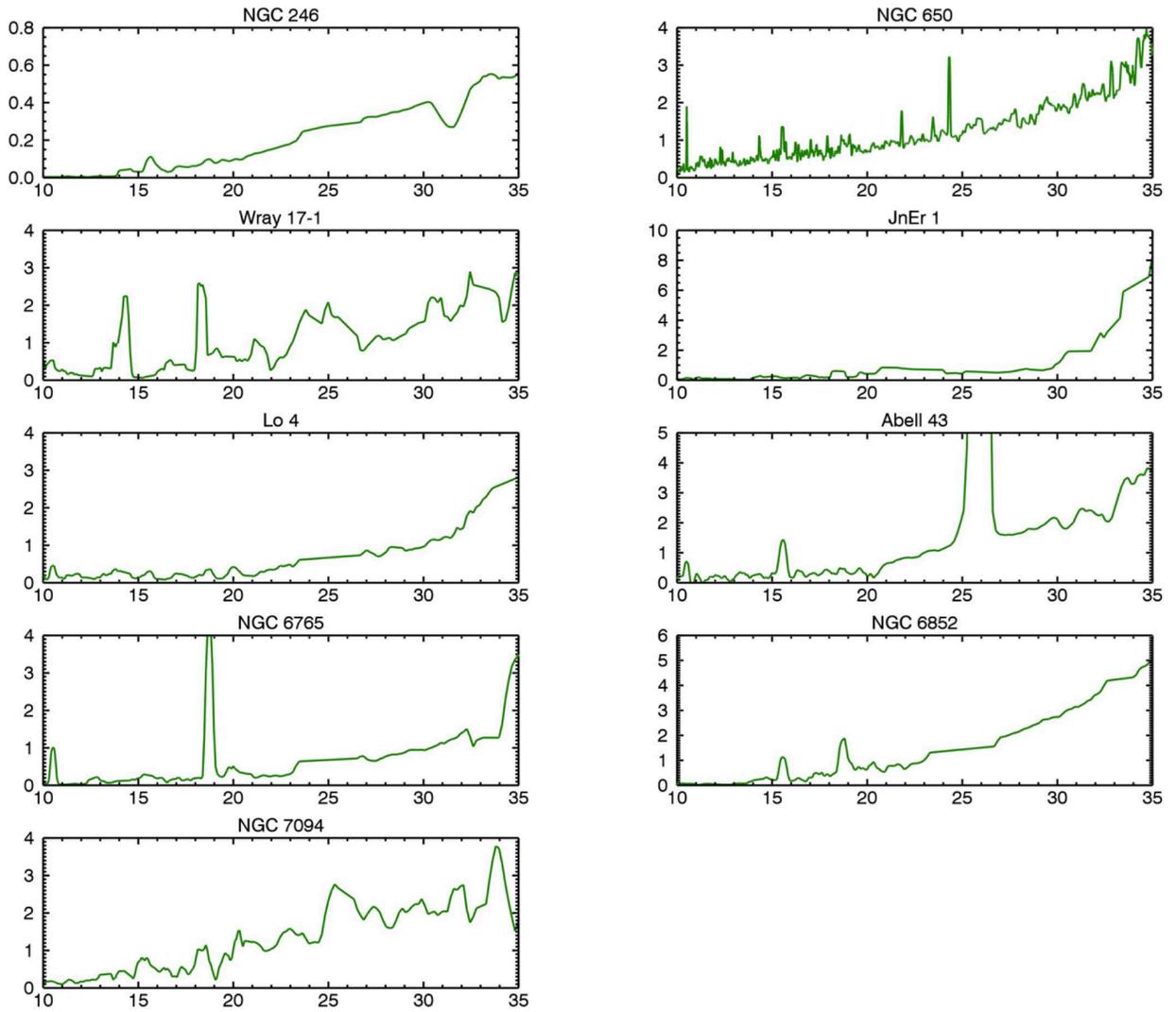}
\end{center}
\caption{IRS spectra for the observed sample of PG 1159 stars plotted as relative flux vs. wavelength, from 10 to 35 \micron. Note that all of the spectra have emission rising toward longer wavelengths, which is associated with dust in the PNe surrounding the target stars. No significant point sources centered on the target stars were detected.}
\end{figure}

\begin{figure}
\figurenum{3} 
\begin{center}
\includegraphics[width=3in,angle=0]{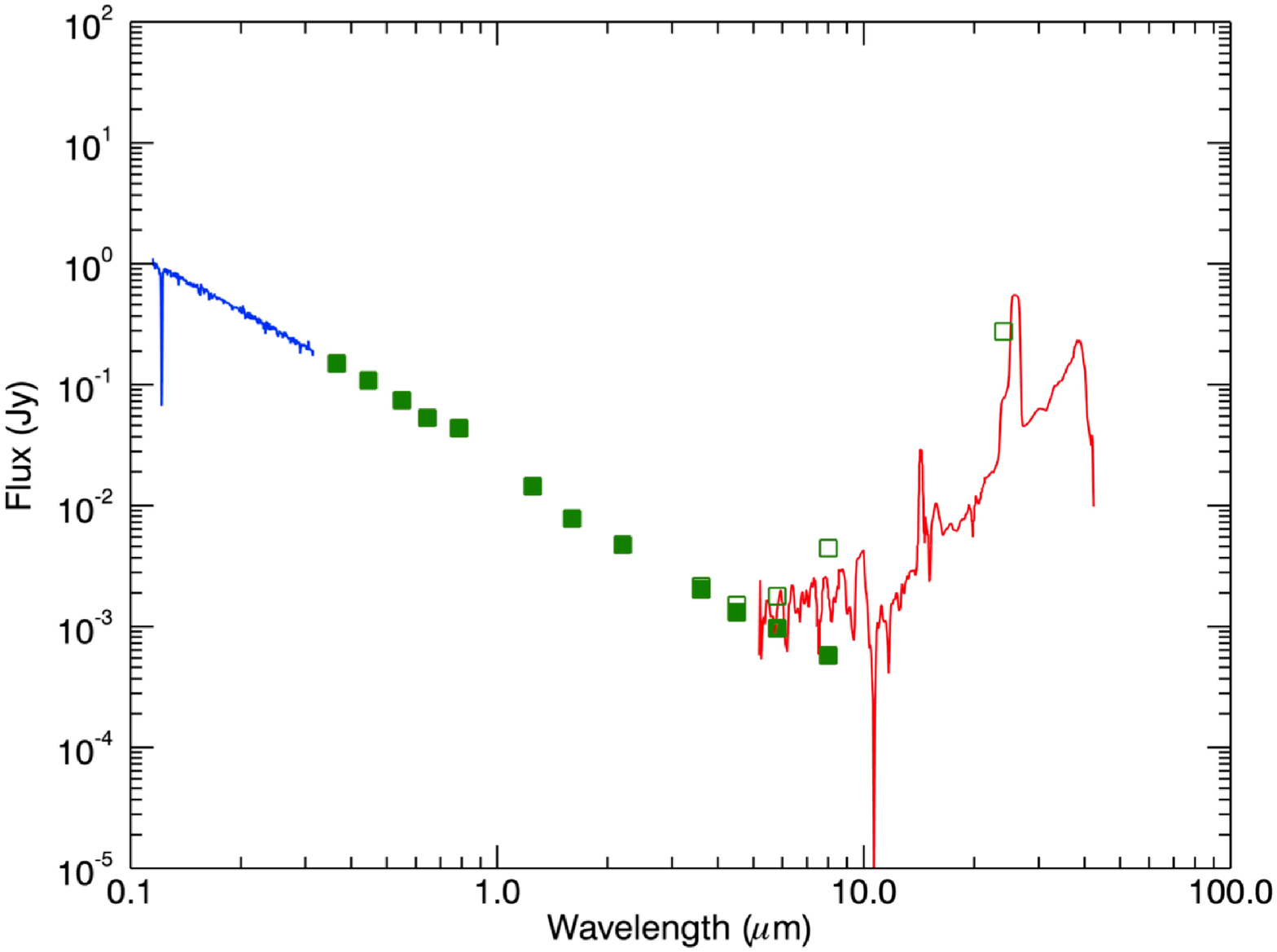}\\
\includegraphics[width=3in,angle=0]{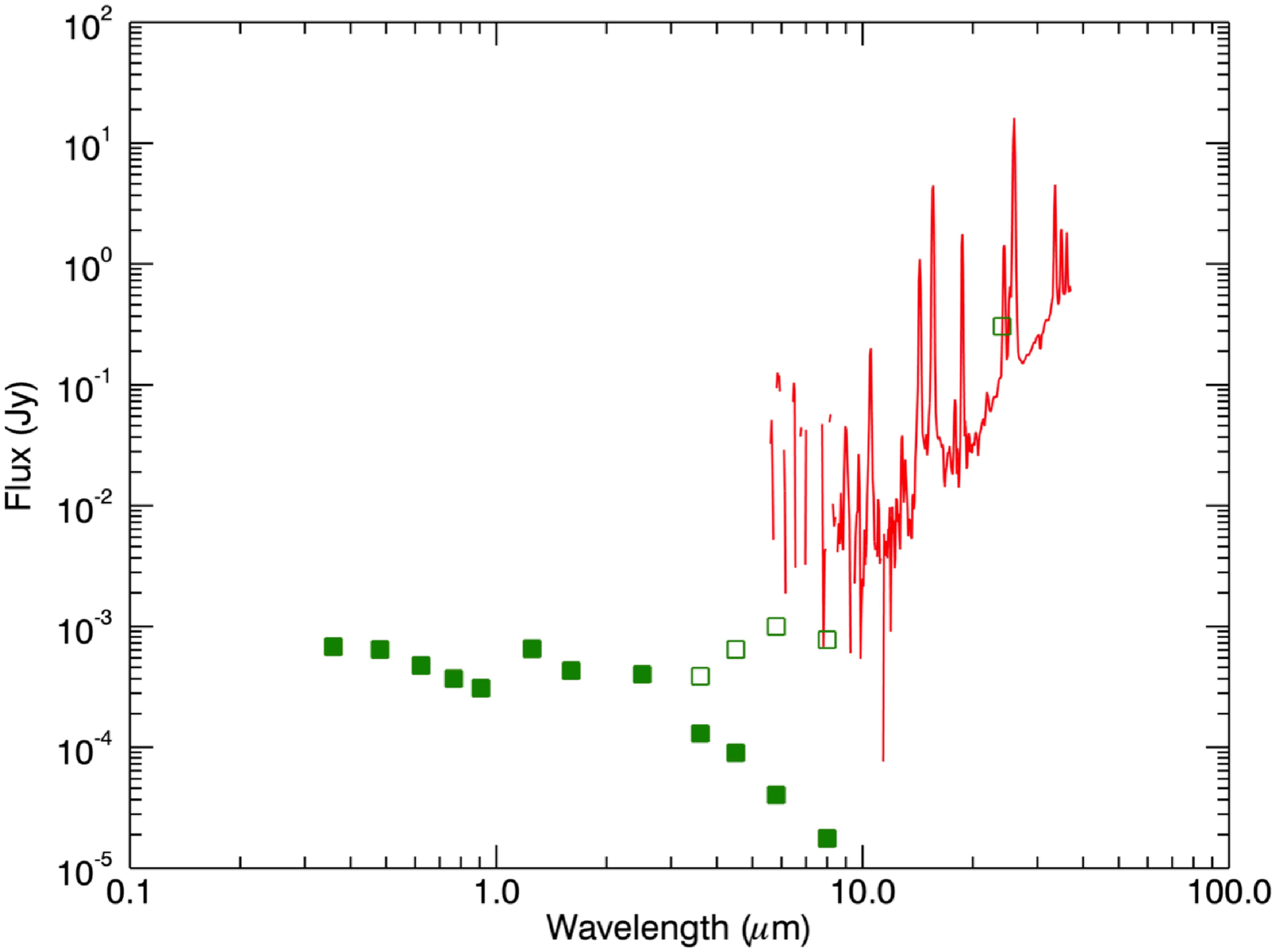}\\
\includegraphics[width=3in,angle=0]{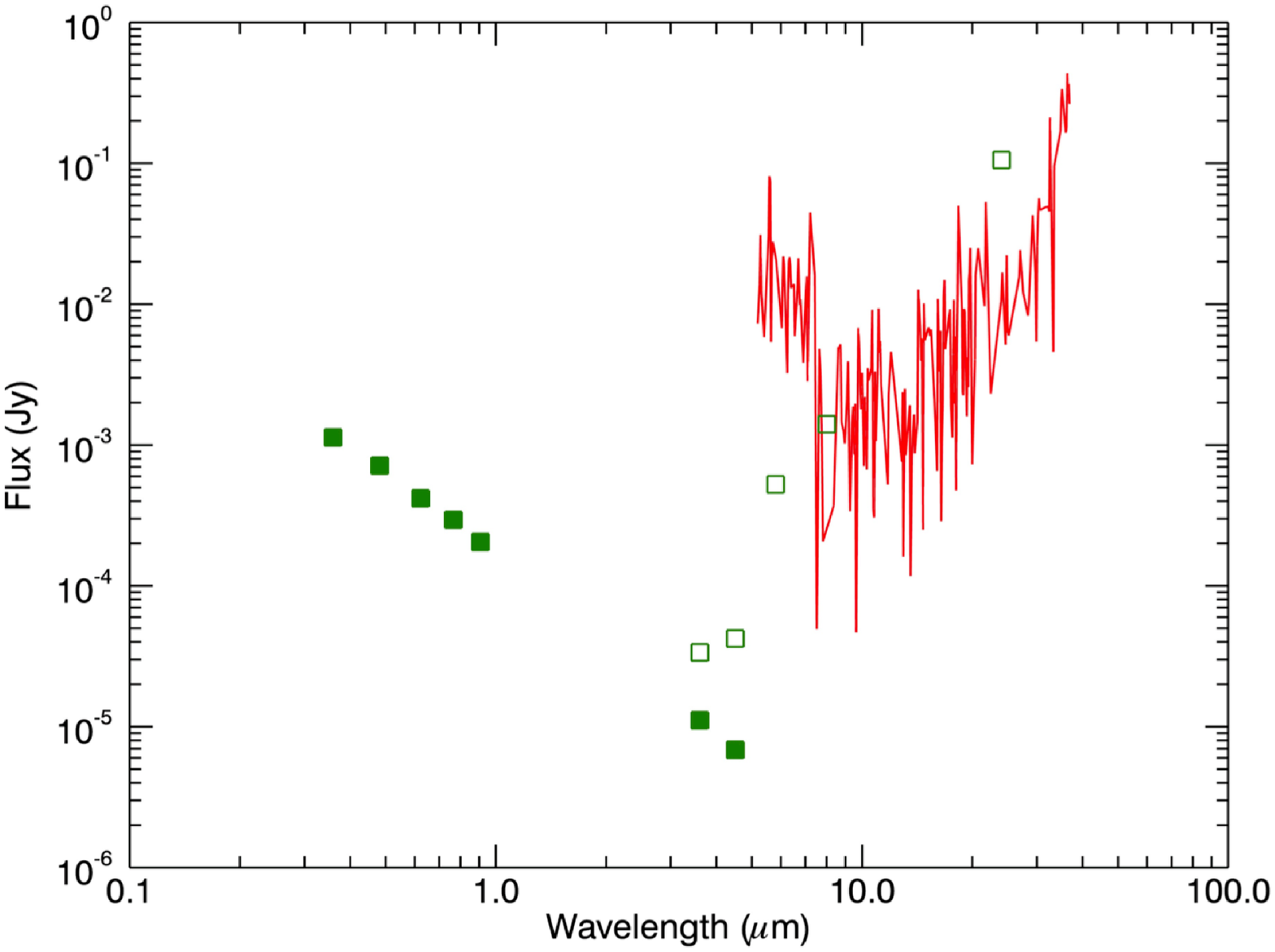}
\end{center}
\caption{SEDs for NGC246 (top panel), NGC 650 (middle panel) and JnEr 1 (bottom panel). The IUE spectrum is in blue and IRS spectra are in red. The solid green squares denote the sky-subtracted photometry in the $U$, $B$, $V$, $R$, $I$, $J$, $H$, and $K$ bands and IRAC  bands at 3.6, 4.5, 5.8 and 8.0 \micron. The open squares denote the photometry without sky subtraction.  The MIPS 24-\micron~point is included. The close companions to NGC 650 are contaminating the photometry in the $J$, $H$ and $K$ bands. }
\end{figure}

\begin{figure}
\figurenum{4} 
\begin{center}
\includegraphics[width=3in,angle=0]{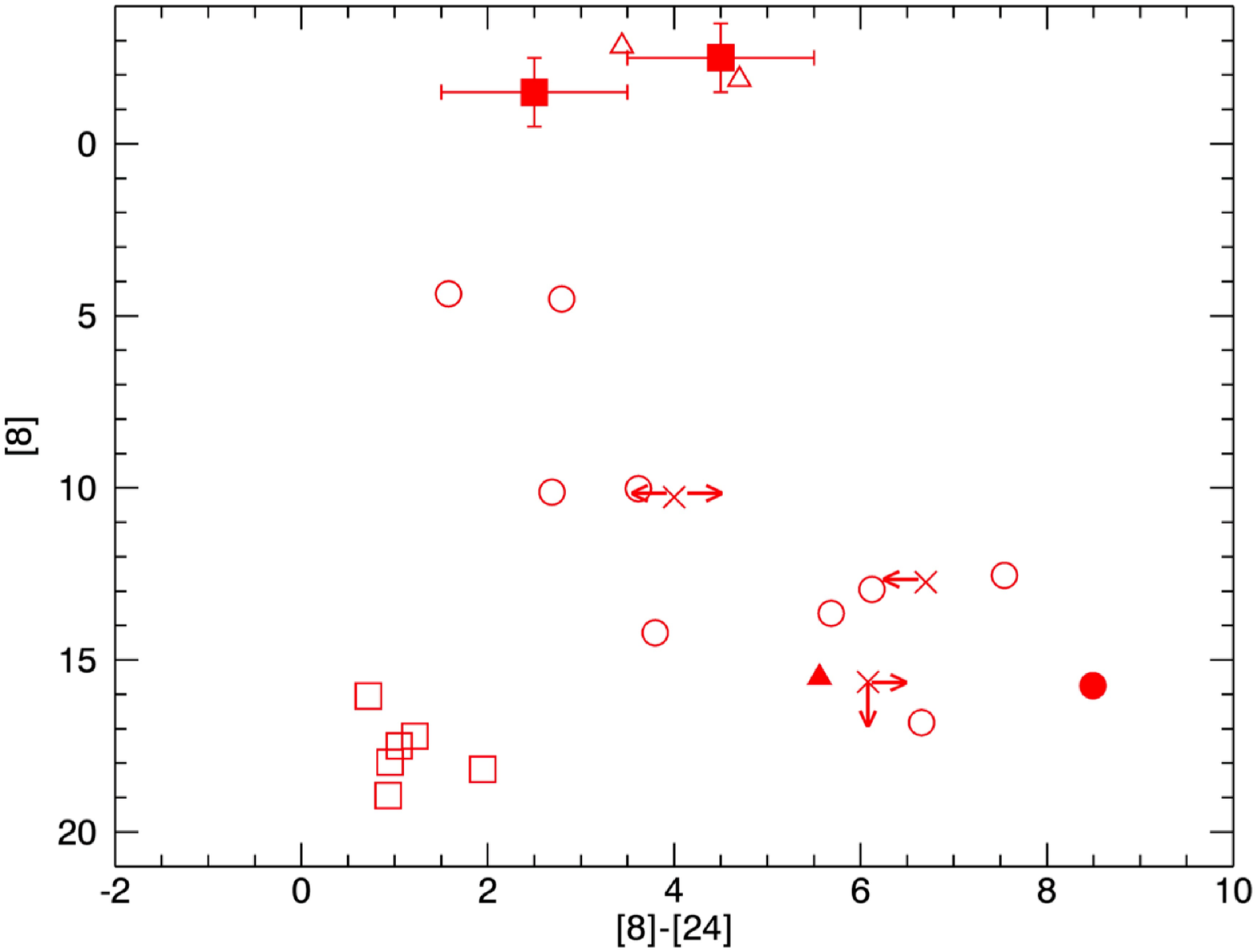}
\end{center}
\caption{
Color-magnitude diagram ([8-24] vs. [8]) plotting IRAC 8 \micron~magnitude against the color, IRAC 8 \micron~magnitude - MIPS 24 \micron~magnitude. 
It shows the average position of 10 LMC post-AGB stars with carbon-rich chemistry and 10 with oxygen-rich chemistry \citep[filled squares;][]{2011MNRAS.411.1597W}, GD 56, GD 133, and four other WDs with disks  \citep[open squares;][]{2007ApJ...663.1285J}, the central stars of the Helix PN \citep[filled circle;][]{2007ApJ...657L..41S}, DeHt~5, a post-RGB star (filled triangle), eight CSPNs and one CSPN mimic (open circles) \citep{2012ApJS..200....3B}, Abell 21  \citep[X + right and down arrows, it was not detected at 8 \micron;][]{2012ApJS..200....3B}, NGC~246  \citep[X + left arrow, it was not detected at 24 \micron;][]{2012ApJS..200....3B}, and NGC 7139 \citep[X + left and right arrows, it was not observed at 24 \micron;][]{2012ApJS..200....3B}. \label{fig:CMD} All fluxes have been scaled to 500 pc and converted to magnitudes.}
\end{figure}

\begin{figure}
\figurenum{5} 
\begin{center}
\includegraphics[width=5in,angle=0]{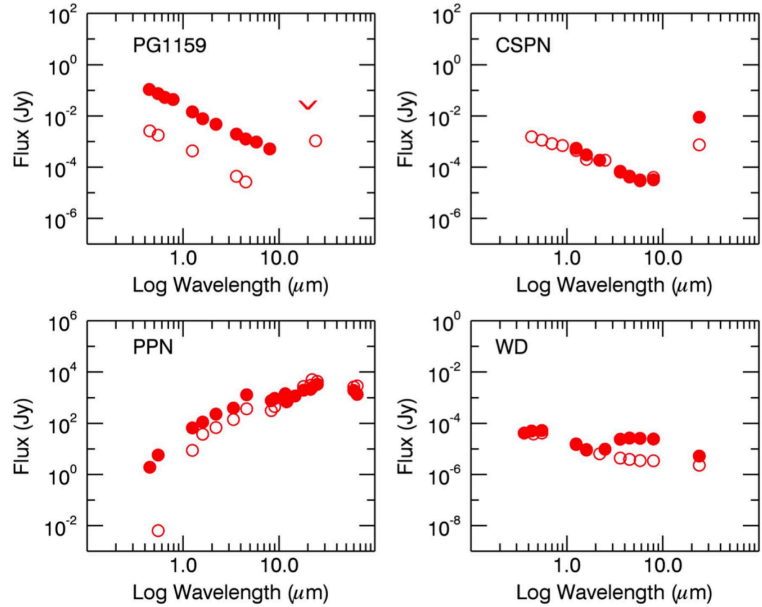}
\end{center}
\caption{Upper Left: SEDs for NGC 246 (filled) and A21 (open). Upper right: SEDs for the central star of the Helix nebula (filled) and DeHt5 (open). Lower left:  SEDs for IRAS 08005-2356 (filled) and IRAS 16279-4757 (open). Lower right: SEDs for GD 56 (filled) and GD 133 (open) All fluxes have been scaled to 500 pc.
\label{fig:SED}}
\end{figure}

\end{document}